\documentclass[preprint,showpacs,preprintnumbers,amsmath,amssymb]{article}

\usepackage{graphicx}
\usepackage{dcolumn}
\usepackage{bm}

\begin{document}

\title{Hamiltonian analysis of $n$-dimensional Palatini gravity with matter}

\author{Muxin Han\footnote{Email\ address:\ hamsyncolor@hotmail.com},
Yongge Ma\footnote{Email\ address:\ mayg@bnu.edu.cn}, You
Ding\footnote{Email\ address:\ ding\_you@hotmail.com},
and Li Qin\footnote{Email\ address:\ qinli051@163.com}\\
\small Department of Physics, Beijing Normal University, Beijing
100875, CHINA}

\date{\today}

\maketitle

\begin{abstract}
We consider the Palatini formalism of gravity with cosmological
constant $\Lambda$ coupled to a scalar field $\phi$ in
$n$-dimensions. The $n$-dimensional Einstein equations with
$\Lambda$ can be derived by the variation of the coupled Palatini
action provided $n>2$. The Hamiltonian analysis of the coupled
action is carried out by a $1+(n-1)$ decomposition of the
spacetime. It turns out that both Palatini action and Hilbert
action lead to the same geometric dynamics in the presence of
$\Lambda$ and $\phi$. While, the $n$-dimensional Palatini action
could not give a connection dynamics formalism directly.


\end{abstract}

{PACS number(s): 04.50.+h, 04.20.Fy}

\section{Introduction}
In Palatini formalism Lorentz connection becomes one of the basic
dynamical variables. This feature causes great interest in the
study of non-perturbative quantum
gravity\cite{Ash}\cite{romano}\cite{AL}, modified gravity
theories\cite{Fla}\cite{Vol} and their cosmological
applications\cite{MW}\cite{Kre}. On the other hand, high
dimensional gravitational theories, such as Kaluza-Klein theory,
are widely
investigated\cite{wesson}\cite{Ma}\cite{yang}\cite{qiang} since
they provide the possibility to unify gravity with gauge fields by
certain higher dimensional geometry. Also the matter and black
holes in high dimensional gravity\cite{a} and the concept of
energy in high dimensional spacetime\cite{c}\cite{b} are fully
investigated. The $n$-dimensional Palatini action, in the case
$n>2$, can reproduce $n$-dimensional vacuum Einstein
equations\cite{ding}. Although the Hamiltonian formalism of 4(or
3)-dimensional Palatini gravity has been fully
studied\cite{Ash}\cite{romano}, the full Hamiltonian analysis of
higher dimensional Palatini formalism is still lacking. One may
ask the question whether one could derive a connection dynamics
formalism from $n$-dimensional Palatini action, as connection
variables are the foundation to apply the technique of loop
quantum gravity.

In this paper, we consider the Palatini formalism of
$n$-dimensional gravity with cosmological constant $\Lambda$
coupled to a Klein-Gordon field $\phi$. A straightforward
calculation show that the coupled Palatini action can still
reproduce $n$-dimensional Einstein's equations even in the
presence of $\Lambda$ and $\phi$ provided $n>2$. We then derive
the corresponding Hamiltonian formulation by carrying out a
$1+(n-1)$ decomposition of the underlying $n$-dimensional
spacetime. It is well known that in the Legendre transform of
4-dimensional Palatini formalism, besides the expected first-class
constraints representing the internal gauge symmetry and spacetime
symmetry, there appear also second-class constraints to account
for the degrees of freedom of the theory. A complicated step in
the Hamiltonian analysis is then to solve those second-class
constraints. While, in higher dimensional Palatini formalism,
another problem arises since the construction of the anticipated
constraints responsible for the degrees of freedom is not so
obvious as that in 4-dimensional case. Let alone to solve the
constraints. We deal with the two troublesome problems by one
trick, namely gauge fixing a unit time-like internal vector
$n^\mu$ and solving the boost part of the Gauss constraint
relative to $n^\mu$. It turns out that both first order Palatini
action and second order Einstein-Hilbert action lead to the same
geometric dynamics in $n$-dimensions $(n>2)$ in the presence of
$\Lambda$ and $\phi$. Thus, as in 4-dimensional case, one could
not obtain connection dynamics directly from the $n$-dimensional
Palatini action.

\section{Lagrangian Formalism}

Consider an $n$-manifold ($n>2$), $M$, on which the basic
dynamical variables in the Palatini framework are $n$-basis
$e_{\mu}^{a}$ and $so(1,n-1)$-valued connection
$\bar{\omega}_{a}^{\ \mu\nu}$(not necessarily torsion-free), where
the Greek indices $\mu,\nu,...$ denote the internal $SO(1,n-1)$
group and the Latin indices $a, b,...$ denote the "spacetime
indices". The internal space is equipped with a Minkowskian metric
$\eta_{\mu\nu}$ (of signature $- + \ldots +$), fixed once for all,
such that the spacetime metric reads:
\begin{displaymath}
g_{ab}=\eta_{\mu\nu}e^{\mu}_{a}e^{\nu}_{b}.
\end{displaymath}
The coupled Palatini action in which we are interested is given
by:
\begin{eqnarray}
S_{p}[e_{\sigma}^{b},\bar{\omega}_{a}^{\mu\nu}, \phi] =
&&\frac{1}{2}\int_{M}d^nx(e)[
e_{\mu}^{a}e_{\nu}^{b}\overline{\Omega}_{ab}^{\ \
\mu\nu}+2\Lambda]\nonumber\\&&-\frac{\alpha_M}{2}\int_{M}d^nx(e)
[\eta^{\mu\nu}e^{a}_{\mu}e^{b}_{\nu}(\bar{\partial}_a\phi)\bar{\partial}_b\phi+m^2\phi^2]
\label{action},
\end{eqnarray}
where $e$ is the square root of the determinant of the $n$-metric
$g_{ab}$, $\Lambda$ is the $n$-dimensional cosmological constant,
$\alpha_M$ is the coupling constant, $\bar{\partial}_a$ is a flat
derivative operator on $M$, and the $so(1,n-1)$-valued curvature
2-form $\overline{\Omega}_{ab}^{\ \ \mu\nu}$ of the connection
$\bar{\omega}_{a}^{\mu\nu}$ reads:
\begin{displaymath}
\overline{\Omega}_{ab}^{\ \ \mu\nu}
=2\mathcal{\overline{D}}_{[a}\bar{\omega}_{b]}^{\ \mu\nu}
\equiv(d\bar{\omega}^{\mu\nu})_{ab}+\bar{\omega}_{a}^{\
\mu\sigma}\wedge\bar{\omega}_{b\sigma}^{\ \ \nu}.
\end{displaymath}
The gravitational field equations are obtained by varying this
action with respect to $e_{\mu}^{a}$ and $\bar{\omega}_{a}^{\
\mu\nu}$. To carry out the variation with respect to the
connection, it is convenient to introduce the unique
(torsion-free) generalized covariant derivative $\bar{\nabla}_{a}$
on both space-time and internal indices determined by the bases
$e_{\mu}^{a}$ via:
\begin{equation}
\bar{\nabla}_{a}e_{\mu}^{b}=\bar{\partial}_a
e_{\mu}^{b}+\bar{\Gamma}^b_{\
ac}e_{\mu}^{c}+\bar{\Gamma}_{a\mu}^{\ \ \nu}e_{\nu}^{b}=0,
\label{basis}
 \end{equation}
where $\bar{\Gamma}^b_{\ ac}$ and $\bar{\Gamma}_{a\mu}^{\ \ \nu}$
are respectively the Levi-Civita connection and spin connection on
$M$. The difference between the $\bar{\omega}_{\ \ a}^{\mu\nu}$
and $\bar{\Gamma}_{a}^{\ \mu\nu}$ is a covariant generalized
tensor field with respect to both internal and spacetime indices
defined by:
\begin{equation}
\bar{C}_{a}^{\ \mu\nu}\equiv\bar{\omega}_{a}^{\
\mu\nu}-\bar{\Gamma}_{a}^{\ \mu\nu}. \label{C1}
\end{equation}
Hence the difference between the curvatures of $\bar{\omega}$ and
$\bar{\Gamma}$ is given by:
\begin{equation}
\overline{\Omega}_{ab}^{\ \ \mu\nu}-\overline{R}_{ab}^{\ \
\mu\nu}=2\bar{\nabla}_{[a}\bar{C}_{b]}^{\ \mu\nu}+2\bar{C}_{[a}^{\
\mu\rho}\bar{C}_{b]\rho}^{\ \ \ \nu},\label{C2}
\end{equation}
where $\overline{R}_{ab}^{\ \ \mu\nu}$ is the curvature 2-form of
$\bar{\nabla}_{a}$. Note that the variation of the action
(\ref{action}) with respect to $\bar{\omega}_{a}^{\ \mu\nu}$
(keeping the basis fixed) is the same as its variation with
respect to $\bar{C}_{a}^{\ \mu\nu}$. Using Eq. (\ref{C2}), the
action (\ref{action}) becomes:
\begin{eqnarray}
S_{p}[e_{\sigma}^b, \bar{C}_{a}^{\ \mu\nu}, \phi]=
&&\frac{1}{2}\int_{M}d^nx(e)[e_{\mu}^{a}e_{\nu}^{b}
(\overline{R}_{ab}^{\ \ \mu\nu}+2\bar{\nabla}_{[a}\bar{C}_{b]}^{\
\mu\nu}+2\bar{C}_{[a}^{\
\mu\rho}\bar{C}_{b]\rho}^{\ \ \ \nu})+2\Lambda]\nonumber\\
&&-\frac{\alpha_M}{2}\int_{M}d^nx(e)[\eta^{\mu\nu}e^{a}_{\mu}e^{b}_{\nu}(\bar{\partial}_a\phi)
\bar{\partial}_b\phi+m^2\phi^2].
\end{eqnarray}
By varying this action with respect to $\bar{C}_{a}^{\ \mu\nu}$,
one obtains:
\begin{equation}
\big(e_{\rho}^{[a}e_{\sigma}^{b]} \delta^{\rho}_{[\mu}
\delta_{\nu]}^{\tau}\big)\bar{C}_{b\tau}^{\ \ \sigma}=0,
\label{last}
\end{equation}
which implies:
\begin{equation}
(n-2)\bar{C}_{a\sigma}^{\ \ \mu}e^a_{\mu}=0.
\end{equation}
 This yields $\bar{C}_{\mu\sigma}^{\ \ \mu}=0$, when $n\neq2$. Using this result, Eq.
(\ref{last}) leads to
\begin{equation}
\bar{C}_{\sigma\mu\ }^{\ \ \nu}=\bar{C}_{(\sigma\mu)}^{\ \ \ \nu}.
\end{equation}
Thus, $\bar{C}_{\sigma\mu\nu}$ is symmetric in its first two
indices. Since $\bar{C}_{\sigma\mu\nu}=\bar{C}_{\sigma[\mu\nu]}$,
we can successively interchange the indices to show
$\bar{C}_{\sigma\mu\nu}=0$. This is the desired result. Thus, the
equation of motion for the connection $\bar{\omega}_{a}^{\
\mu\nu}$ is simply that it equals $\bar{\Gamma}_{a}^{\ \mu\nu}$.
Thus the connection $\bar{\omega}_{a}^{\ \mu\nu}$ is completely
determined by the bases. By carrying out the variation of action
(\ref{action}) with respect to the bases, one obtains:
\begin{eqnarray}
e_{\mu}^c \overline{\Omega}_{cb}^{\ \
\mu\nu}-\frac{1}{2}\overline{\Omega}_{cd}^{\ \
\sigma\rho}e_{\rho}^c e_{\sigma}^d e^{\nu}_b-\Lambda
e^{\nu}_b=\alpha_M
\big[\eta^{\mu\nu}e^c_{\mu}(\bar{\partial}_b\phi)\bar{\partial}_c\phi\nonumber\\
-\frac{1}{2}e^{\nu}_b\big(\eta^{\mu\sigma}e^{c}_{\mu}e^{d}_{\sigma}(\bar{\partial}_c\phi)
\bar{\partial}_d\phi+m^2\phi^2\big)\big].
\label{Einstein}
\end{eqnarray}
Using the fact that $\overline{\Omega}_{ab}^{\ \
\mu\nu}=\overline{R}_{ab}^{\ \ \mu\nu}$ and the curvature 2-form
of $\bar{\nabla}_a$ is related to its space-time curvature by
$\overline{R}_{ab\mu}^{\ \ \ \nu}=\overline{R}_{abc}^{\ \ \
d}e_{\mu}^{c}e^{\nu}_{d}$ and multiplying Eq. (\ref{Einstein}) by
$e_{\nu a}$, it follows that the Einstein equations holds. It is
obvious that the variation of action (\ref{action}) with respect
to $\phi$ will still give the Klein-Gordon equation.

\section{Hamiltonian Analysis}
To carry out the Hamiltonian analysis of action (\ref{action}),
suppose the spacetime $M$ is topologically $\Sigma\times R$ for
some $(n-1)$-manifold $\Sigma$. We introduce a foliation and a
time-evolution vector field $t^a$ in $M$, where $t^a$ can be
decomposed with respect to the unit normal vector $n^a$ of
$\Sigma$ as:
\begin{equation}
t^a=Nn^a+N^a,
\end{equation}
where $N$ is called the \textit{lapse function} and $N^a$ called
the \textit{shift vector}\cite{wald}\cite{Liang}. Denote
$S_p=S_G+S_{KG}$. Then the action of gravity and matter can be
respectively decomposed as:
\begin{eqnarray}
S_G=&&\frac{1}{2}\int_{\Sigma\times R}d^nx(E)[NE^a_\mu
E^b_\nu\Omega_{ab}^{\ \
\mu\nu}+2n_{[\mu}E^a_{\nu]}\mathcal{D}_a\bar{\omega}^{\
\mu\nu}_{t}-2n_{[\mu}E^a_{\nu]}\dot{\omega}^{\
\mu\nu}_{a}\nonumber\\&&+2N^an_{[\mu}E^b_{\nu]}\Omega_{ab}^{\ \
\mu\nu}+2N\Lambda],\\
S_{KG}=&&-\frac{\alpha_M}{2}\int_{\Sigma\times
R}d^nx(E)N[\eta^{\mu\nu}E^{a}_{\mu}E^{b}_{\nu}(\partial_a\phi)\partial_b\phi
-\frac{1}{N^2}(\dot{\phi}-N^a\partial_a\phi)^2\nonumber\\&&+m^2\phi^2]\label{1},
\end{eqnarray}
where $E^a_\mu\equiv e^b_\mu q^a_b=e^b_\mu(g^a_b+n^an_b)$, $E$ is
the square root of the determinant of the spatial metric $q_{ab}$,
$\mathcal{D}_a$ is the spatial $SO(1,n-1)$ generalized covariant
derivative operator reduced from $\mathcal{\overline{D}}_a$ and
corresponds to a $so(1,n-1)$-valued spatial connection 1-form
$\omega_a^{\ \mu\nu}\equiv q_a^b\bar{\omega}_b^{\ \mu\nu}$,
$\Omega_{ab}^{\ \ \mu\nu}\equiv
q_a^cq_b^d\overline{\Omega}_{cd}^{\ \ \mu\nu}$ is the
corresponding spatial $so(1,n-1)$-valued curvature 2-form,
$\dot{\omega}_a^{\ \mu\nu}$ is the Lie derivative of
${\omega}_a^{\ \mu\nu}$ with respect to $t^a$ (treating the
internal indices as scalars), and $\partial_a$ is the derivative
operator on $\Sigma$ reduced from $\bar{\partial}_a$. The internal
normal vector is defined as $n_\mu\equiv n_ae^a_\mu$, from which
one has $\eta^{\mu\nu}E^a_\mu n_\nu=0$. By a gauge fixing
$n_\mu\equiv(1, 0, 0...)$, one can split the internal indices
$\mu, \nu, \sigma,...$ into $0, i, j,...$ . Note that the gauge
fixing put no restriction on our real dynamics.

Let $\underline{N}\equiv N/E$ be the densitized lapse scalar of
weight $-1$ and $\widetilde{E}^a_i\equiv(E)E^a_i$ the densitized
spatial basis of weight $1$. The action of gravitational field can
then be decomposed as:
\begin{eqnarray}
S_G=&&\int_{\Sigma\times R}d^nx[\underline{N}\widetilde{E}_i^a
\widetilde{E}_j^b(D_{[a}\omega_{b]}^{\
ij}+K^i_{[a}K^j_{b]})-\bar{\omega}_t^{\
i0}D_a\widetilde{E}^a_i\nonumber\\&&+\bar{\omega}_t^{\
ij}\widetilde{E}^a_{[i}K_{|a|j]}+\widetilde{E}^a_i\dot{K}^i_a
-2N^a\widetilde{E}^b_jD_{[a}K^j_{b]}+(E)^2\underline{N}\Lambda],
\end{eqnarray}
where $K^j_a\equiv\omega_a^{\ j0}$, and $D_a$ is the $SO(n-1)$
generalized covariant derivative operator with respect to
$\omega_a^{\ ij}$. The unique torsion-free $SO(n-1)$ generalized
covariant derivative operator annihilating $E^a_i$ is defined as:
\begin{equation}
\nabla_{a}E_{i}^{b}=\partial_a E_{i}^{b}+\Gamma^b_{\
ac}E_{i}^{c}+\Gamma_{ai}^{\ \ j}E_{j}^{b}=0,
\end{equation}
where $\Gamma^b_{\ ac}$ and $\Gamma_{ai}^{\ \ j}$ are respectively
the Levi-Civita connection and the spin connection on $\Sigma$.
Let $C_a^{\ ij}$ be the difference between $\omega_a^{\ ij}$ and
$\Gamma_{a}^{\ ij}$, i.e.,
\begin{equation}
\omega_a^{\ ij}=\Gamma_a^{\ ij}+C_a^{\ ij}.
\end{equation}
Then the constraint equation with respect to the Lagrangian
multiplier $\bar{\omega}_t^{\ i0}$ reads:
\begin{equation}
\frac{\delta S_p}{\delta\bar{\omega}_t^{\
i0}}=D_a\widetilde{E}^a_i=\nabla_a\widetilde{E}^a_i+C_a^{\
ij}\widetilde{E}^a_j=(E)C_j^{\ ij}=0,\label{C}
\end{equation}
which means that $C_j^{\ ij}=-C_j^{\ ji}=0$ in the reduced phase
space. So in the reduced phase space determined by Eq. (\ref{C}),
the action of gravitational field reads
\begin{eqnarray}
S_G=&&\int_{\Sigma\times
R}d^nx[\frac{1}{2}\underline{N}\widetilde{E}_i^a
\widetilde{E}_j^bR_{ab}^{\ \
ij}-\frac{1}{2}(E)^2\underline{N}C_{j\ k}^{\ i}C_i^{\
kj}+\underline{N}\widetilde{E}_i^{[a}
\widetilde{E}_j^{b]}K^i_aK^j_b\nonumber\\&&+\bar{\omega}_t^{\
ij}\widetilde{E}^a_{[i}K_{|a|j]}+\widetilde{E}^a_i\dot{K}^i_a-2N^a\widetilde{E}^b_j\nabla_{[a}K_{b]}^j+N^aC_a^{\
ij}\widetilde{E}^b_jK_{bi}\nonumber\\&&+(E)^2\underline{N}\Lambda],
\end{eqnarray}
where $R_{ab}^{\ \ ij}\equiv2\nabla_{[a}\Gamma_{b]}^{\ ij}$ is the
Riemann curvature 2-form compatible with the spatial basis
$E^a_i$. Then the variation of $S_p$ respect to $C_j^{\ ik}$
gives:
\begin{equation}
\frac{\delta S_p}{\delta C_j^{\
ik}}=-\frac{1}{2}(E)^2\underline{N}(C_{[ik]}^{\ \ j}+C_{[k\ i]}^{\
j})+N^j\widetilde{E}^b_{[k}K_{|b|i]}=0.\label{a}
\end{equation}
While, the constraint equation determined by the Lagrangian
multiplier $\bar{\omega}_t^{\ ij}$ reads:
\begin{equation}
\frac{\delta S_p}{\delta \bar{\omega}_t^{\
ij}}=\widetilde{E}^a_{[i}K_{|a|j]}=0.\label{b}
\end{equation}
Since $C_i^{\ jk}=C_i^{\ [jk]}$, by substituting Eq. (\ref{b})
into Eq. (\ref{a}), one gets:
\begin{equation}
C_{[ik]}^{\ \ j}=0,
\end{equation}
which leads to $C_{ijk}=0$ by using the same trick in the previous
section. Hence, the action of gravitational field can be reduced
to:
\begin{eqnarray}
S_G=&&\int_{\Sigma\times
R}d^nx[\widetilde{E}^a_i\dot{K}^i_a+\underline{N}\big(\frac{1}{2}(E)^2(R+2\Lambda)+\widetilde{E}_i^{[a}
\widetilde{E}_j^{b]}K^i_aK^j_b\big)+\bar{\omega}_t^{\
ij}\widetilde{E}^a_{[i}K_{|a|j]}\nonumber\\&&-2N^a\widetilde{E}^b_j\nabla_{[a}K_{b]}^j].\label{c}
\end{eqnarray}

On the other hand, the action of the scalar field can be written
via the internal gauge fixing as:
\begin{eqnarray}
S_{KG}=&&-\frac{\alpha_M}{2}\int_{\Sigma\times
R}d^nx[\delta^{ij}\underline{N}\widetilde{E}^{a}_{i}\widetilde{E}^{b}_{j}(\partial_a\phi)\partial_b\phi
-\frac{1}{\underline{N}}(\dot{\phi}-N^a\partial_a\phi)^2\nonumber\\&&+(E)^2\underline{N}m^2\phi^2].
\end{eqnarray}
The canonical momentum conjugating to $\phi$ reads:
\begin{eqnarray}
\widetilde{\pi}=\frac{\partial\cal
L}{\partial\dot{\phi}}=\frac{\alpha_M}{\underline{N}}(\dot{\phi}-N^a\partial_a\phi).
\end{eqnarray}
Hence the scalar field action can be expressed as:
\begin{eqnarray}
S_{KG}=&&\int_{\Sigma\times
R}d^nx[\widetilde{\pi}\dot{\phi}-\underline{N}\big(\frac{\alpha_M}{2}\delta^{ij}\widetilde{E}^{a}
_{i}\widetilde{E}^{b}_{j}(\partial_a\phi)\partial_b\phi+\frac{1}{2\alpha_M}\widetilde{\pi}^2
+\frac{\alpha_M}{2}(E)^2m^2\phi^2\big)\nonumber\\&&-N^a\widetilde{\pi}\partial_a\phi].\label{d}
\end{eqnarray}
Thus, combining Eqs. (\ref{c}) and (\ref{d}), the total
Hamiltonian and all constraint equations of $n$-dimensional
Palatini gravity with $\Lambda$ coupled to $\phi$ can be
summarized as:
\begin{eqnarray}
H_{tot}=&&-\int_{\Sigma\times R}d^{n-1}x[\bar{\omega}_t^{ij}\mathcal{G}_{ij}+\underline{N}\mathcal{H}
+N^a\mathcal{V}_a]\\
\mathcal{G}_{ij}=&&\widetilde{E}^a_{[i}K_{|a|j]},\label{e}\\
\mathcal{H}=&&[\frac{1}{2}(E)^2(R+2\Lambda)+\widetilde{E}_i^{[a}\widetilde{E}_j^{b]}K^i_aK^j_b]\nonumber\\
&&-[\frac{\alpha_M}{2}\delta^{ij}\widetilde{E}^{a}_{i}\widetilde{E}^{b}_{j}\partial_a\phi\partial_b\phi
+\frac{1}{2\alpha_M}\widetilde{\pi}^2+\frac{\alpha_M}{2}(E)^2m^2\phi^2],\label{f}
\\
\mathcal{V}_a=&&-2\widetilde{E}^b_j\nabla_{[a}K_{b]}^j-\widetilde{\pi}\partial_a\phi.\label{g}
\end{eqnarray}
They have the same form as those in the ADM Hamiltonian formalism.
Hence the constraints (\ref{e})-(\ref{g}) also comprise a
first-class system.

In conclusion, for arbitrary $n>2$ dimensional spacetime, the
Palatini action and the Einstein-Hilbert action lead to the same
classical dynamics in the presence of cosmological constant and
Klein-Gordon field. Thus, as an alternative approach, one may
study the dynamics of higher dimensional gravitation with matter
fields, such as the brane world theory,  in Palatini formalism as
well. While, the $n$-dimensional Palatini action could not give a
connection dynamics formalism directly.

\section*{ Acknowledgments}

This work is  supported in part by NSFC (10205002) and YSRF for
ROCS, SEM. Muxin Han, You Ding and Li Qin  would also like to
acknowledge support from Undergraduate Research Foundation of BNU.

\end{document}